\documentclass{aa}

\usepackage{graphicx}
\usepackage{txfonts}
\usepackage{tabularx}
\usepackage{amssymb}
\usepackage{amsmath}
\usepackage{multirow}
\usepackage{color,soul}
\graphicspath{{figures/}}
\usepackage{tablefootnote}
\usepackage{footnote}
\makesavenoteenv{tabular}
\makesavenoteenv{table}
\usepackage{natbib}
\bibpunct{(}{)}{;}{a}{}{,} 
\makeatletter
\newcommand\textlist[4]{%
  \let\last@item\relax
  \let\last@sep\relax
  \@for\@ii:=#4\do{%
    \ifx\last@item\relax\else
      \ifx\last@sep\relax
        \def\last@sep{#2}%
      \else#1\fi
      #3{\last@item}%
    \fi
    \let\last@item\@ii
  }%
  \ifx\last@item\relax\else
    \last@sep#3{\last@item}%
  \fi
}
\makeatother

\newcommand{\citett}{\textlist{; }{ and }{\citet}}
\begin{document}
 
   \title{Deriving the Hubble constant using \textit{Planck} and \textit{XMM-Newton} observations of galaxy clusters}


   \author{Arpine Kozmanyan     
          \inst{1,2}
          \and
          Hervé Bourdin
          \inst{1,2}
          \and
          Pasquale Mazzotta
          \inst{1,2}
          \and
          Elena Rasia
          \inst{3}
          \and
          Mauro Sereno
          \inst{4,5}
          }

   \institute{Dipartimento di Fisica, Universit\`a di Roma ``Tor~Vergata'', via della Ricerca Scientifica 1, I-00133, Roma, Italy
         \and
         Sezione INFN Roma~2, via della Ricerca Scientifica 1, I-00133, Roma, Italy
         \and
         INAF, Osservatorio Astronomico di Trieste, via Tiepolo 11, I-34131, Trieste, Italy
         \and
         INAF - Osservatorio di Astrofisica e Scienza dello Spazio di Bologna,
via Piero Gobetti 93/3, I-40129 Bologna, Italy
                 \and
                 Dipartimento di Fisica e Astronomia, Alma Mater - Universita di Bologna, via Piero 
Gobetti 93/2, I-40129 Bologna, Italy
             }
             
   \date{Received July 16, 2018}

    \abstract
    {The possibility of determining the value of the Hubble constant using observations of galaxy clusters in X-ray and microwave wavelengths through the Sunyaev Zel’dovich (SZ) effect has long been known. Previous measurements have been plagued by relatively large errors in the observational data and severe biases induced, for example, by cluster triaxiality and clumpiness. The advent of \textit{Planck} allows us to map the Compton parameter $y$,  that is, the amplitude of the SZ effect, with unprecedented accuracy at intermediate cluster-centric radii, which in turn allows performing a detailed spatially resolved comparison with X-ray measurements.  Given such higher quality observational data, we developed a Bayesian approach that combines informed priors on the physics of the intracluster medium obtained from hydrodynamical simulations of massive clusters with measurement uncertainties.\par 
We applied our method to a sample of 61 galaxy clusters with redshifts up to $z < 0.5$ observed with \textit{Planck} and \textit{XMM-Newton}  and find $H_0=67 \pm 3$km s$^{-1}$ Mpc$^{-1}$.}


        \maketitle
%

\section{Introduction}
\label{introduction}

The X-ray radiation from galaxy clusters and the spectral distortion of the cosmic microwave background (CMB) radiation by inverse Compton scattering of CMB photons (Sunyaev-Zel'dovich effect, SZ) are both due to the electrons in the intracluster medium (ICM)~\citep{sarazin1988, Sunyaev1970}. The amplitudes of the two effects have different dependence on the density of the electrons in the ICM. The two effects can be jointly used to break the degeneracy existing in the amplitudes of the signals between the cluster electron density and the angular diameter distance to the cluster. Some early works by~\citett{cowie, gunn, silk, cavaliere78} proposed this method to constrain cosmological parameters such as the Hubble constant, the deceleration parameter, or the flatness of the universe. A well-known advantage of the method is that it does not depend on any secondary cosmic scales and is based on very simple principles.\par
Early estimations by~\citet{birkinshaw, reese2000, patel2000, mason2001, reese2002, sereno2003, udomprasert2004, schmidt2004, jones2005} and others relied on data from ground-based low-frequency ($10$ - $150$ GHz) radio interferometers detecting the decrement side of the thermal SZ (tSZ) distortion. As an example,~\citet{reese2003} combined SZ measurements from the \textup{\textit{Ryle telescope} (\textit{RT}), \textit{Owens Valley Radio Observatory} (\textit{OVRO})} and the \textit{Berkeley-Illinois-Maryland Association} (\textit{BIMA}) observatories and X-ray measurements from the \textit{ROSAT} satellite for $26$ clusters within redshift $z \leq 0.78$ and found a value of $H_0=61 \pm 3$(stat.)$ \pm 18$(sys.) km s$^{-1}$ Mpc$^{-1}$ for a flat $\Lambda$CDM cosmology with $\Omega_m=0.3$ and $\Omega_{\Lambda}=0.7$. Using SZ measurements from the \textit{OVRO}, \textit{BIMA}, and X-ray measurements from the \textit{Chandra} observatory for $38$ clusters in the redshift range $0.14 \leq z \leq 0.89$,~\citet{Bonamente2006} estimated $H_0={76.9^{+3.9}_{-3.4} }$(stat.)$^{+10.0} _{-8.0}$(sys.) km s$^{-1}$ Mpc$^{-1}$ for the same cosmological model. A slightly higher accuracy was achieved by~\citet{schmidt2004}, who used only $\text{three}$ regular clusters at redshifts equal to $0.088$, $0.2523$, and $0.451$ and found $H_0=68 \pm 8,$ again for the same cosmological model. The authors claimed that they were able to achieve an improved accuracy using only regular systems as the systematic errors are negligible with respect to the statistical ones.\par       
Recent estimates of the Hubble constant from CMB anisotropies by~\textit{Planck Collaboration}, ($H_0=66.93 \pm 0.62$ km s$^{-1}$ Mpc$^{-1}$,~\citealt{Planck1}) have reached a precision that is hard to compete with. Nevertheless, because of the known discrepancy of this value with the value derived by~\citet{ries2018} using Cepheid-calibrated type Ia supernovae (SNIa) ($H_0=73.48 \pm 1.66$ km s$^{-1}$ Mpc$^{-1}$ or $H_0=73.52 \pm 1.62$ km s$^{-1}$ Mpc$^{-1}$ from~\citealt{ries20181}), it is key to have external and independent confirmations. The development of alternative methods will allow us to advance our understanding of the issue and of the potential sources of discrepancy, be they of cosmological or of systematic origin. For this reason, a number of new and promising approaches are brought forward, such as the use of water masers~\citep{reid2013, gao2016}, or lensed multiple images of quasars~\citep{serenograv, wong, bonvin} or SNe~\citep{grillo2018}. \par

In line with this thought, and in view of newly available data from the \textit{Planck} space satellite, we revisit the method of determining $H_0$ using observations of galaxy clusters. The frequency coverage of the CMB spectrum offered by the high-frequency instrument of \textit{Planck} (100 - 857 GHz) allows mapping the Compton parameter $y$, that is, the amplitude of the SZ effect, with an unprecedented accuracy at intermediate cluster-centric radii~\citep{Bourdin}. It thus permits us to perform precise, spatially resolved comparisons with X-ray measurements. Given these measurement accuracies, the limiting factor of the method now becomes our knowledge of the ICM physics and geometry, which motivates the introduction of priors from hydrodynamical simulations of massive clusters. We developed a Bayesian approach that combines such priors with measurement uncertainties. In this paper we discuss this method and its application to a sample of 61 moderately distant galaxy clusters observed with \textit{Planck} and \textit{XMM-Newton}. The method might also allow a future determination of the helium abundance in cluster gas. \par
\vspace{0.3cm}
The paper is structured as follows. In Sections~\ref{sample} and~\ref{simulations} we present the observed and simulated samples, respectively. In Section~\ref{method} we describe our method. We present the collective characterisation of the most prominent bias sources by making use of simulated galaxy clusters. We applied this information to correct real observations in order to place constraints on cosmological parameters. In Section~\ref{result} we present our results, and in Section~\ref{discuss} we compare them with previous measurements and predict precisions that will be possible using this method.

\section{Observed sample}
\label{sample}
Our sample is derived from the PSZ2 catalogue of SZ-selected clusters by the \textit{Planck} mission. The set of clusters used for this work is almost identical to the set used in~\cite{Planck2} for the X-ray - SZ scaling relations. The original set counted $62$ clusters in total, but one of them ($ZwCl 1215+0400$) was later excluded from the second \textit{Planck} catalogue of SZ sources. In this work, we removed this cluster and used a total of $61$ galaxy clusters in the redshift range $0<z<0.5$. The mass range of the clusters in the sample is $2.6\times 10^{14} M_{\odot} \leq M_{500} \leq 1.8\times 10^{15} M_{\odot} $.\par
Studies of cluster populations selected in SZ and X-ray surveys indicate that  SZ-selected samples could be a fair representation of the general population of clusters in the universe~\citep{rosetti2015, rosetti2017, sereno2017, felipe}. Unlike flux-limited X-ray selected cluster populations that seem to preferentially include dynamically relaxed and cool-core clusters, the SZ-selected clusters do not exhibit such preference~\citep{rosetti2015, rosetti2017, felipe}. The clusters selected via SZ appear to be unbiased representatives of the overall cluster population since their density profile and concentrations are consistent with standard predictions of $\Lambda$CDM cosmology~\citep{sereno2017}.\par
As described in~\cite{Planck2}, because the selection of our sample combines both SZ and X-ray criteria, we cannot fully claim that the dataset used in our analysis is representative or complete. It represents a large sample of clusters observed homogeneously with multi-frequency milimetric and X-ray observations of suitable angular resolution, however, allowing us to keep the statistical errors to the minimum.
In the near future, large cluster projects within the Heritage program\footnote{e.g. witnessing the culmination of structure formation in the universe galaxies, groups of galaxies, clusters of galaxies, and superclusters, PI M. Arnaud and S. Ettori.} of \textit{XMM-Newton} will resolve this specific issue and provide access to large samples of mass-selected clusters. \par

\section{Simulated sample}
\label{simulations}

We used the hydrodynamical simulations of galaxy clusters presented in~\cite{rasia}. They were carried out with the improved version of the TreePM-smooth-particle-hydrodynamics code GADGET-3~\citep{springel} introduced in~\cite{beck2015}. The runs considered uniform time-dependent ultraviolet (UV) background and a radiative cooling that is metallicity dependent~\citep{wiersma}. Star formation and evolution were modelled in a sub-resolution fashion from a multi-phase gas as in~\cite{spring2003}. Metals were produced by SNIa, SNII, and asymptotic-giant-branch stars as in~\cite{tornatore}. Galactic winds of velocity 350 km s$^{-1}$ mimicked the kinetic feedback by SN. The active galactic nucleus (AGN) feedback followed the~\cite{steinborn} model, where both mechanical outflows and radiation were evaluated separately. Their combined effect was implemented in terms of thermal energy. Only cold accretion onto the black holes was considered, which was computed by multiplying the Bondi rate by a boost factor $\alpha=100$. The accretion was Eddington limited. Further details can be found in~\citett{rasia, planelles, biffi2017}.\par
\begin{figure}
\centering
\includegraphics[width=7cm]{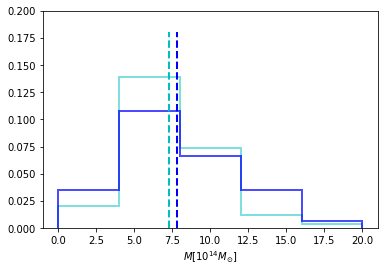}
\caption{Normalised mass distributions of observed and simulated samples. In cyan we plot the mass distribution in the observed sample. The dashed line represents the median of the distribution at $7.3\times 10^{14} M_{\odot}$. In blue we show the mass distribution in the simulated set. The dashed line represents the median at $7.8\times 10^{14} M_{\odot}$.
}
\label{fig:mass_comp}
\end{figure}
The cosmological model in the simulation assumes a flat $\Lambda$CDM cosmology with $H_0=72$ km s$^{-1}$ Mpc$^{-1}$ and $\Omega_m=0.24$ and a fraction of hydrogen mass $X=0.76$.\par
These simulations agree largely in their properties with those exhibited by samples of observed clusters. For instance, a comparison of their entropy profiles with the profiles measured by \cite{pratt2007} shows a remarkable agreement~\citep{rasia}. The pressure profiles from~\cite{planelles} are in line with the observational results by \citett{arnaud, planckpress, Sayers, Sun, Bourdin}. They find general agreement between simulated and observed sets within $0.2 \leq r/R_{500} \leq 1$. They also study the properties of the clumpiness of this set and show that the 3D median radial distribution of the clumping factor at $z=0$ is in reasonable agreement with observations by~\cite{eckert} - the largest observational sample studied for clumpiness so far. Finally,~\cite{biffi2017} compared radial profiles of iron abundance with the observations of~\cite{leccardi} and found agreement within the dispersion of the simulated profiles.\par
We here analyse clusters with masses in the range $2.6\times 10^{14} M_{\odot} \leq M_{500} \leq 1.8\times 10^{15} M_{\odot} $ at different redshifts. Namely, $(i)$ 26 galaxy clusters at $z=0$, $(ii)$ 25 clusters at $z=0.25$, and $(iii)$ 21 clusters at $z=0.5$. This subsample was chosen from the overall sample of~\cite{rasia} to ensure similar mass ranges for the observed and simulated clusters. In Fig.\ref{fig:mass_comp} we show the mass distributions of the two samples. The two distributions have similar shape and medians: $7.3\times 10^{14} M_{\odot}$ and $7.8\times 10^{14} M_{\odot}$ for the observed and simulated sets, respectively. This also demonstrates that the balance of low- and high-mass clusters in the two sets is comparable. \par
In order to increase the sample size, we took three perpendicular projections of each cluster and calculated them as three different clusters. This gave us a total sample size of $216$ clusters. It is important to note that the different projections and redshift snapshots of the same clusters are not completely independent. In Appendix~\ref{app-d} we present arguments that ensure that the overall distribution created in this way does not introduce additional biases due to correlation between the sample constituents.\par
The redshift ranges and mass distribution of the simulated sample are thus similar to those of the observed sample. This and the above-mentioned proximity of the simulated cluster properties to the properties of observed clusters indicates that the simulated clusters provide a fair representation of the observed set of clusters used in this analysis. \par

\section{Method}
\label{method}
In this section we describe the procedure we followed to estimate the value of $H_0$ using SZ and X-ray observations. This can be subdivided into three stages: i) joint deprojection of the ICM profiles given the SZ and X-ray observables, ii) characterisation of biases of non-cosmological origin, and iii) estimating the value of the Hubble constant.\par
\subsection{Joint deprojection of the ICM profiles}
\label{profiles}
To estimate the 3D electron number density $n_e$, temperature $kT,$ and pressure $P_e$ profiles, we used the fitting procedure of~\cite{Bourdin}, which we summarise below.\par
\subsubsection{Derivation of $n_e(r)$}

First, X-ray data were used to constrain the 3D $n_e(r)$ profile and to provide an initial approximation of the $kT(r)$ profile. We assumed spherical symmetry and modelled the observable quantities with the analytical profiles suggested by~\cite{vikh}. More specifically, for the electron number density, we used
\begin{equation}
\begin{aligned}
n_pn_e(r)=\frac{n_0^2(r/r_c)^{-\alpha'}}{[1+(r/r_c)^2]^{3 \beta_1-\alpha'/2}} \frac{1}{[1+(r/r_s)^{\gamma}]^{\epsilon/\gamma}}\\
 +\frac{n_{02}^2}{[1+(r/r_{c2})^2]^{3 \beta_2}},
\end{aligned}
\label{eqn:number}
\end{equation}
where $r_c$ and $r_{c2}$ are the characteristic radii of $\beta$-like profiles with slopes $\beta_1$ and $\beta_2$, with a power-law cusp modification parametrised with the index $\alpha'$; $n_0$ and $n_{02}$ are the normalisations of the two components at the centre; and $r_s$ is the characteristic radius in the outer steeper regions of the profile with slope $\epsilon$. \par
For the temperature we used
\begin{equation}
kT(r)=T_0\frac{x+T_{min}/T_0}{x+1} \frac{(r/r_t)^{-a}}{(1+(r/r_t)^b)^{c/b}},
\label{eqn:temp}
\end{equation}
where $x \equiv (r/r_{cool})^{a_{cool}}$, and $r_{cool}$ describes the scale of the central cooling region with slope $a_{cool}$ and normalisation $T_{min}$; $r_t$, $a$, $b$, and $c$ describe the size and profile slopes outside the cooling region; and $T_0$ is the overall normalisation of the profile.\par
\vspace{0.1cm}
After integrating these 3D models along the line of sight (LOS), 
\begin{equation}
\Sigma_x(r)=\frac{1}{4 \pi (1+z)^3} \int [n_p n_e](r) \Lambda(T,Z) dl,
\label{eqn:surf}
\end{equation}
\begin{equation}
kT_X(r)=\frac{\int w kT(r) dl}{\int wdl} \text{, with } w=n_e^2/T^{3/4},
\label{eqn:mazz}
\end{equation}
we fit them jointly to the observed projected X-ray surface brightness $\Sigma^{obs}_x(r)$ and temperature $kT^{obs}_X(r)$ profiles obtained from \textit{XMM-Newton} (see~\citealt{Bourdin} for details).\par 
In Eq.~\ref{eqn:surf} the surface brightness has units cnt s$^{-1}$ cm$^{-2}$ arcmin$^{-2}$. If energy units are used, such as erg s$^{-1}$ cm$^{-2}$ arcmin$^{-2}$, then the factor $1+z$ in the denominator would be to the fourth power, that is, $\Sigma_x(r)=\frac{1}{4 \pi (1+z)^4} \int [n_p n_e](r) \Lambda(T,Z) dl$. In Eq.~\ref{eqn:mazz} the temperature weighting from~\cite{mazz} is used.\par
Modelling the factor $\Lambda(T,Z)$ assumes dependence on the metallicity of the gas and weak dependence on temperature. The first was modelled assuming a redshifted, Galactic-hydrogen-absorbed spectral energy distribution (SED) of hot gas with bremsstrahlung continuum and metal emission lines as tabulated in the Astrophysical Plasma Emission Code (APEC,~\citealt{smith}). We adopted the solar composition of metal abundances tabulated by~\cite{Grevesse} and a constant normalisation of 0.3. The fit of the $kT(r)$ in this part of the procedure took care of the weak temperature dependence of the factor $\Lambda(T,Z)$.\par
Finally, we derived the electron number density profile $n_e(r)=\sqrt{n_e/n_p \times [n_pn_e](r)}$ assuming $n_p/n_e=0.852$, which corresponds to a helium abundance of $Y=0.2527$ for the metal abundances cited above.\par

\subsubsection{Derivation of $P(r)$ and $kT(r)$}

In the second step, we jointly fit the X-ray projected temperature $kT^{obs}_X(r)$ and SZ $y^{obs}(r)$ signal profiles. $kT^{obs}_X(r)$ was extracted from the \textit{XMM-Newton}, while $y^{obs}(r)$ was extracted from the six \textit{Planck} HFI maps (see~\citealt{Bourdin} for details).\par
To model the SZ signal, we used the analytical gNFW pressure profile from~\cite{nagai}:
\begin{equation}
P_e(r)=\frac{P_0}{(c_{500}x)^{\gamma}(1+(c_{500}x)^{\alpha})^{(\beta-\gamma)/\alpha}},
\label{eqn:press}
\end{equation}
with $x \equiv r/r_{500}$ and $r_{500}$ being defined as the radius of the cluster within which the mean density of the cluster is $500$ times higher than the critical density of the Universe at the clusters redshift. $P_0$ is the overall normalisation of the profile, $c_{500}$ is the concentration with respect to $r_{500}$ , and $\alpha$, $\beta,$ and $\gamma$ are the slopes at the inner, intermediate, and outer regions of the profile, respectively.\par
By fixing the $n_e(r)$ profile to the form obtained in the previous step, we created a template for the temperature as
\begin{equation}
kT(r)=\eta_T \times P_e(r)/n_e(r).
\label{eqn:temp-sz}
\end{equation} 
We then integrated Eqs.~\ref{eqn:press} and~\ref{eqn:temp-sz} along the LOS using
\begin{equation}
y(r)= \frac{\sigma_T}{m_e c^2} \int P_e(r) \times dl
\end{equation}
for the pressure profile and Eq.~\ref{eqn:mazz} for the temperature, and we fit them jointly to the observed projected $y^{obs}(r)$ and $T^{obs}_X(r)$ profiles. In addition to estimating the 3D pressure profile $P_e(r)$, this procedure also returns the value of the normalisation parameter $\eta_T$, which reflects the discrepancy between the measurement of pressure profile using only X-ray or SZ observables.\par 
In the ideal case of spherical symmetry with no clumpiness, we expect $\eta_T=1$. In the realistic case, instead, $\eta_T$ is expected to be different from 1, and its departure depends on different aspects such as the assumptions of the underlying cosmological model and/or some ICM distribution properties (elongation, orientation, clumpiness, etc.). \par
As shown in Appendix~\ref{app-a}, $\eta_T$ has a simple dependence from the main properties that can be divided into two terms:
\begin{equation}
\eta_T = \mathcal{C} \times \mathcal{B}. 
\label{eqn:eta-fin}
\end{equation}
The first term depends only on quantities that are directly related to the cosmological parameters (such as $H_0$, or $Y$). It is defined as
\begin{equation}
\mathcal{C}= \left(\frac{\bar{D}_{a}}{D_a} \right)^{1/2} \times \left(\frac{{n_p/n_e}}{{\bar{n}_p/\bar{n}_e}}\right)^{1/2} \times \left(\frac{1+ 4 \frac{n_{He}}{n_p}}{1+ 4 \frac{\bar{n}_{He}}{\bar{n}_p}}\right)^{1/2},
\label{eqn:eta-fin2}
\end{equation}
where $D_a$ is the angular diameter distance, $n_p/n_e$ is the ratio of the hydrogen to electron number density, and $\frac{n_{He}}{n_p}$ the ratio of helium to hydrogen number density. The latter two factors both depend on the helium abundance $Y$ in the cluster gas (see Appendix~\ref{app-a}).\par 
The second term contains everything else that is not directly related to the cosmological model or the helium abundance. It can be parametrised as
\begin{equation}
\mathcal{B}=b_n \frac{C^{1/2}_{\rho}}{e^{1/2}_{LOS}}, 
\label{eqn:eta-fin1}
\end{equation}
where $e_{LOS}$ is a factor that accounts for the cluster asphericity, $C_{\rho}=\frac{<\rho^2>}{<\rho>^2}$ accounts for the cluster clumpiness, and the factor $b_n$ denotes any other bias that could arise from our profile modelling and/or the fitting procedure.\par
In the previous two formulas, the non-bar and bar notations of parameters refer to their true values and to the values assumed in the data analysis above, respectively. More specifically, we used a $\Lambda$CDM cosmological model with $\bar{H}_0=70$ km s$^{-1}$ Mpc$^{-1}$, $\bar{\Omega}_m=0.3$ and $\bar{\Omega}_{\Lambda}=0.7$, $\bar{n}_p/\bar{n}_e=0.852,$ and $\frac{\bar{n}_{He}}{\bar{n}_p}=0.0851$. For a fully ionised medium, the latter correspond to an assumption of a helium abundance of $\bar{Y}=0.2527$. We refer to Appendix~\ref{app-a} for a detailed derivation of the Eqs.~\ref{eqn:eta-fin},~\ref{eqn:eta-fin2},~and \ref{eqn:eta-fin1} above. To constrain the cosmological quantities, we need to characterise the contribution of the $\mathcal{B}$ term, which is addressed in the next subsection.\par

\subsection{Characterisation of $\mathcal{B}$}
\label{prior}
To characterise the $\mathcal{B}$ term in Eq.~\ref{eqn:eta-fin}, we adopted a simple procedure based on the use of our set of cosmological hydrodynamic simulations that have a clumpiness level comparable to that of real clusters~\citett{eckert, planelles}.\par
We assumed that the cosmological parameters are known, and we fixed them to the values we used for the simulated sample, that is, a flat $\Lambda$CDM cosmology with $H_0=72$ km s$^{-1}$ Mpc$^{-1}$ and $\Omega_m=0.24$, a ratio of hydrogen to electrons $n_p/n_e=0.864,$ and a number density of helium to hydrogen $n_{He}/n_p=0.0789$ from the fraction of hydrogen mass $X=0.76$ (see Section~\ref{simulations}). These assumptions guarantee that for the simulated sample $\mathcal{C}=1$ (see Eq.~\ref{eqn:eta-fin2}). Then, for every simulated cluster, we estimated the parameter $\eta_T$ applying the exact same procedure used for the real clusters. \par
In this respect, to imitate the ``observed" 2D quantities, we projected the simulated cluster properties by integrating along $10$ Mpc in the direction of LOS:
\begin{equation}
\Sigma_x(r_{min}, r_{max})=\frac{\sum m_i \rho_i}{A_{ring[r_{min}, r_{max}]}/\pi},
\end{equation}
\begin{equation}
T_X(r_{min}, r_{max})=\frac{\sum T_i w_i V_i}{\sum T_i V_i}, 
\end{equation}
and
\begin{equation}
y(r_{min}, r_{max})=\frac{\sigma_T}{m_e c^2} \frac{\sum P_i V_i}{A_{ring[r_{min}, r_{max}]}},
\end{equation}
with the sum extending to all particles within a cylinder of radius $[r_{min}, r_{max}]$ and height $10$Mpc; $m_i$, $\rho_i$, $T_i$, $P_i$ , and $V_i$ being the mass, density, temperature, pressure, and the volume of the $i$-th particle, respectively, $w_i$ is the spectroscopic-like weight equal to $w_i=\rho_i^2/T_i^{3/4}$ (\citealt{mazz}), and $A_{ring[r_{min}, r_{max}]}=\pi (r_{max}^2-r_{min}^2)$ is the surface area of the cylinders base. \par
\begin{figure}
\centering
\includegraphics[width=7cm]{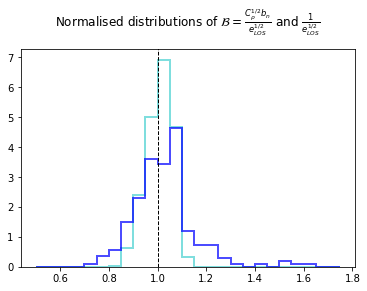}
\caption{In cyan we plot the distribution of the quantity $\frac{1}{e^{1/2}_{LOS}}$ that was calculated semi-analytically. In blue we show the distribution of the quantity $\mathcal{B} \equiv \frac{C^{1/2}_{\rho}}{e^{1/2}_{LOS}} b_n$ , which we used as a prior in our Bayesian estimation of the Hubble constant. The dashed line at value $1$ is shown for reference.
}
\label{fig:priors}
\end{figure}

In Fig.\ref{fig:priors} we show as a blue histogram the $\eta_T$ distribution resulting from this procedure. Assuming that the simulated clusters accurately approximate the real ones in terms of i) shape, the gas shape at $r_{500}$ does not strongly depend on the ICM physics and mostly follows the total potential of the cluster (see ~\citealt{lau2011} and~\citealt{kawahara2010}), and ii) clumpiness level (see~\citealt{planelles} for comparison), the blue histogram gives the intrinsic distribution of the $\mathcal{B}$ term.\par

We expect that asphericity will play a major role. To distinguish its effect from that of clumpiness, we derived the distribution of $e_{LOS}^{-0.5}$ using the semi-analytical approach of~\cite{sereno2017}, which for completeness we also report in Appendix~\ref{app-e}. The result is overlaid as a cyan histogram in Fig.~\ref{fig:priors}. Comparing the two distributions in Fig.~\ref{fig:priors}, we see that asphericity is indeed important, but it is not the only player, and other terms significantly contribute to the $\mathcal{B}$ distribution. The contribution of the remaining components ($C_{\rho}$ and $b_n$) in $\mathcal{B}$ results in mainly a larger dispersion and a more skewed distribution. Thus, asphericity alone would fail to describe the complete effect of the non-cosmological components, and considering the complete distribution $\mathcal{B}$ is crucial to treat them correctly.\par

\subsection{Derivation of the value of $H_0$}
\label{final}
Our  approach in this subsection ensures that all contributions in $\mathcal{B}$ are considered and are corrected for when deriving cosmological parameters. More precisely, we use the distribution of $\mathcal{B}$ as a prior for the non-cosmological bias term in the derivation of $H_0$ given our data $\eta_T$.\par

The model contains $H_0$, $Y_p$, and $\Omega_m$ as parameters in common among all clusters (see Eq.~\ref{eqn:eta-fin}). In addition, we have a nuisance parameter $\mathcal{B}_i$ for each cluster $i$. For $N$ data points (sample of $N$ clusters), we have $N+3$ parameters in the model. The posterior distribution of $H_0$ is
\begin{multline}
P(H_0) = \int \mathcal{L}(\{\eta_T^{(i)}\} | H_0, \Omega_M, Y, \{\mathcal{B}_i\})
\times p(\{\mathcal{B}_i\}) p(H_0)p(\Omega_M)p(Y)\times \\ \times d\Omega_M dYd\mathcal{B}_1d\mathcal{B}_2...d\mathcal{B}_N,
\end{multline}
where we have denoted $\{\mathcal{B}_i\}$ the set of $N$ nuisance parameters $\mathcal{B}_i$. The first factor in the integral is the likelihood. The next four factors in the integral are  the prior probabilities of the combined bias terms $\{\mathcal{B}_i\}$ ($p(\{\mathcal{B}_i\})$) and of the parameters $H_0$, $\Omega_M$, $Y$ - $p(H_0)$, $p(\Omega_M)$, and $p(Y)$. We include the parameters $\Omega_M$ and $Y$ in this equation, since we would like to include the effect of the  uncertainty on these parameters in the final value of $H_0$.\par
Assuming uncorrelated cluster measurements, we can write $\mathcal{L}(\{\eta_T^{(i)}\} | H_0, \Omega_M, Y, \{\mathcal{B}_i\}) = \prod \mathcal{L}(\eta_T^{(i)} | H_0, \Omega_M, Y, \mathcal{B}_i)$, where now by $\mathcal{B}_i$ and $\eta_T^{(i)}$ we denote the particular parameter $\mathcal{B}$ and measurement $\eta_T$ for the $i$th cluster.\par
Simultaneously, since cluster shapes and clumpiness can be expected not to be correlated, we can also write $p(\{\mathcal{B}_i\})=\prod p(\mathcal{B}_i)$.\par
Then the complete form of the posterior probability of $H_0$ is\begin{multline}
P(H_0)= \int \prod_i \mathcal{L}(\eta_T^{(i)} | H_0,  \Omega_M, Y, \mathcal{B}_i)\times p(\mathcal{B}_i) p(H_0) p(\Omega_M)p(Y) \times \\ \times d\Omega_M dYd\mathcal{B}_1d\mathcal{B}_2...d\mathcal{B}_N,
\end{multline}
where the shape of $p(\mathcal{B}_i)$ is defined by the distribution derived above. The prior distribution $p(H_0)$ is taken to be uniform between $50$ and $100$, and null otherwise. The prior distribution $p(\Omega_M)$ is also taken to be uniform between $0.25$ and $0.35$, and null otherwise. Finally, the prior distribution $p(Y)$ is taken to be uniform between $0.24$ and $0.25$, and null otherwise. The measured distribution of $\eta_T^{(i)}$ is shown in Fig.~\ref{fig:etadata}.\par
The actual application of the distribution shown in Fig.~\ref{fig:priors} (limited to the range $0.7<\mathcal{B}_i<1.65$) as a prior distribution of $\mathcal{B}_i$ requires extending the distribution over the entire range of values over which the sampling of $\mathcal{B}_i$ is done. In order to accomplish this, we approximated the tails of the distribution with a Gaussian distribution (see Appendix~\ref{app-c} for details and plot). The left side of the distribution is in agreement with a Gaussian tail (see Fig.~\ref{fig:pdf_prior}), hence its extension with a Gaussian can be assumed to describe the real distribution over the ranges $0<\mathcal{B}_i<0.7$ well. The right tail of the distribution is flatter, however, and cannot be described by a Gaussian. This indicates that in the future, a more complete distribution could help to better reconstruct the right tail of $p(\mathcal{B}_i)$. We plan to use a larger set of simulated clusters for this in a forthcoming work. For the Markov chain Monte Carlo (MCMC) sampling we used the PyMC python open source MCMC sampler~\citep{pymc}.\par

\section{Results}
\label{result}
\begin{figure}
\centering
\includegraphics[width=8cm]{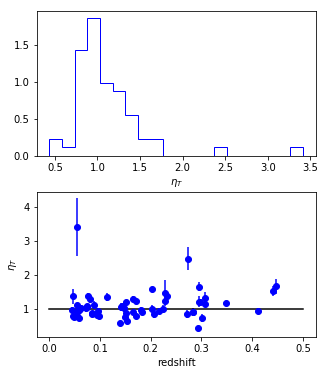}
\caption{Distribution of values of $\eta_T$ derived for a sample of $61$ clusters using the method described in Section~\ref{method} (top). The same values as a function of redshift (bottom). }
\label{fig:etadata}
\end{figure}

The procedure described in Subsection~\ref{profiles} returns 61 $\eta_T$ values, one for each  observed cluster. In Fig.~\ref{fig:etadata} we show the distribution of these values as well as the same values as a function of redshift. As shown in the previous section, the dispersion of the distribution is due to the intrinsic scatter $\mathcal{B}$ shown in Fig.~\ref{fig:priors} convolved with the measurement error. The main cosmological information comes from the shift relative to $p(\mathcal{B})$ that depends on the true value of $H_0$. \par
\begin{figure}
\centering
\includegraphics[width=8cm]{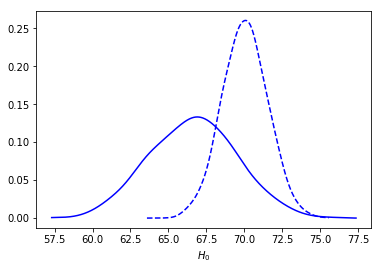}
\caption{Posterior distributions of $H_0$ for a flat $\Lambda$CDM universe. The dashed line is the value of $H_0$ when the bias correction is not applied. The solid line represents the posterior for $H_0$ with the correction for the biases included.}
\label{fig:helium}
\end{figure}

We applied the procedure described in Subsection~\ref{final} to constrain $H_0$ with the observed  $\eta_T$ distribution. We derived the posterior distribution of $H_0$ in two cases. At first, we ignored the intrinsic non-cosmological biases accounted for in the $\mathcal{B}$ parameter. This means that for this test, we assumed spherical symmetry and regularity for galaxy clusters, or in other words, $\mathcal{B}_i \equiv 1$. In Fig.~\ref{fig:helium} we show the corresponding posterior distribution of $H_0$ with a dashed line.\par
Subsequently, we derived the posterior probability of $H_0$ by also taking into account the bias terms $\mathcal{B}_i$ and assuming the prior $p(\mathcal{B}_i)$ obtained from simulations (see Subsection~\ref{prior}). The corresponding posterior distribution is overlaid in Fig.~\ref{fig:helium} as a continuous line.\par
The comparison of the posterior probabilities in Fig.~\ref{fig:helium} suggests that the inclusion of $p(\mathcal{B}_i)$ leads to i) a broader distribution, and ii) a shift of the mean value. The first arises because we included additional uncertainties. The shift, instead, is caused by the fact that the intrinsic distribution of $\mathcal{B}$ is non-symmetrical (see Fig.~\ref{fig:priors}). \par
We finally report the estimated values within $1\sigma$ significance level error: $H_0=70 \pm 1.5$ km s$^{-1}$ Mpc$^{-1}$, when we ignore the intrinsic biases and $H_0=67\pm3$ km s$^{-1}$ Mpc$^{-1}$, when we correctly account for it. \par

\section{Discussion and conclusions}
\label{discuss}
We introduced a new approach for measuring the value of the Hubble constant using SZ and X-ray observations of clusters of galaxies. The method allows for a simultaneous treatment of the statistical and systematic errors. In this section we compare and discuss our result with previous measurements made using SZ and X-ray observations of clusters as well as other estimations performed with different probes. 
\subsection{Comparison with other measurements using SZ and X-ray data}
\begin{table*}
\label{tab:values}
\small{
\centering
\begin{tabular}{ c c c c c c c c c}
Article & Number & redshift & $\Omega_m$, $\Omega_{\Lambda}$ & value & SZ data source & X-ray data source\\\hline

\cite{reese2000} & $2$ & $0.55$ & $0.3$, $0.7$ & ${63^{+12}_{-9}}^{+21}_{-21}$ & OVRO, BIMA & ROSAT\\

\cite{patel2000} & $1$ & $0.322$ & $0.3$, $0.7$ & ${52.2^{+11.4}_{-11.9}}^{+18.5}_{-17.7}$ & OVRO, BIMA, MMT\footnotemark & ROSAT, ASCA\footnotemark\\

\cite{mason2001} & $7$ 
 & $<0.1$ & $0.3$, $0.7$ & ${66^{+14}_{-11}}^{+15}_{-15}$ & OVRO & ROSAT\\
 
\cite{grainge} & $1$ & $0.143$ & $1$, $0$ & ${57^{+23}_{-16}}$ & RT & ROSAT, ASCA \\

\cite{reese2002} & $18$ & $0.14-0.78$ & $0.3$, $0.7$ & ${60^{+4}_{-4}}^{+13}_{-18}$ & OVRO, BIMA & ROSAT\\

\cite{saunders} & $1$ & $0.217$ & $0.3$, $0.7$ & ${85^{+20}_{-17}}$ & RT & ROSAT, ASCA\\

\cite{reese2003} & $26$ & $0-0.78$ & $0.3$, $0.7$ & ${61 \pm 3 \pm 18}$ & RT, OVRO, BIMA & ROSAT\\

\cite{gemma2003} & $1$ & $0.0231$ & $0.27$, $0.73$ & ${84 \pm 26}$ & OVRO, WMAP\footnotemark, MITO\footnotemark & ROSAT\\

\cite{udomprasert2004} & $7$ & $<0.1$ & $0.3$, $0.7$ & ${67^{+30}_{-18}}^{+15}_{-6}$ & CBI & ROSAT, ASCA, BeppoSAX\footnotemark \\

\cite{schmidt2004} & $3$ & $0.09-0.45$ & $0.3$, $0.7$ & ${69 \pm 8}$ & various & Chandra\\

\cite{jones2005} & $5$ & $0.14-0.3$ & $0.3$, $0.7$ & ${66^{+11}_{-10}}^{+9}_{-8}$ & RT & ROSAT, ASCA\\

\cite{Bonamente2006} & $38$ & $0.14-0.89$ & $0.3$, $0.7$ & & OVRO, BIMA & Chandra\\
 &  \multicolumn{3}{l}{double $\beta$-model with HSE}  & ${76.9^{+3.9}_{-3.4}}^{+10.0}_{-8.0}$ & &\\

 & \multicolumn{2}{l}{isothermal $\beta$-model} &   & ${73.7^{+4.6}_{-3.8}}^{+9.5}_{-7.6}$ &  & \\

 &  \multicolumn{3}{l}{isothermal $\beta$-model with excised core} & ${77.6^{+4.8}_{-4.3}}^{+10.1}_{-8.2}$ &  & \\

\end{tabular}
\caption{Non-comprehensive list of measurements of $H_0$ made since 2000 using observations of X-ray and Sunyaev-Zel'dovich galaxy clusters.}
}
\end{table*}

\begin{figure}
\centering
\includegraphics[width=8cm]{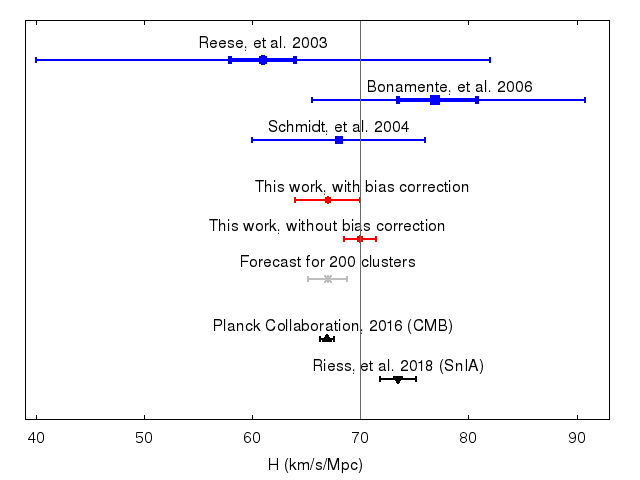}
\caption{Comparison of the result in this work (in red) with previous
measurements using clusters as a probe (in blue) and with current most precise measurements of $H_0$ using other probes (in black). In grey we show our forecast for a future calculation of $H_0$ with the described method and a sample of $200$ clusters.}
\label{fig:dist}
\end{figure}

In Table~\ref{tab:values} we report a non-comprehensive list of measurements of $H_0$ from SZ and X-ray observations of clusters. In addition to the instruments used for the observations, we also report the sizes of the samples and the corresponding estimates of $H_0$. These values are all consistent within the statistical plus systematic errors. This may also be due to relatively large total errors (which are at least $14\%$). For comparison purposes, in Fig.~\ref{fig:dist} we report the measurement obtained using the two largest cluster samples (i.e.~\citealt{reese2003} and~\citealt{Bonamente2006}) together with ours. We also add the measurement of~\cite{schmidt2004}, which, although they used a sample composed of only $\text{three}$ clusters, returns the smallest uncertainty. According to the authors, the use of regular clusters significantly reduces the possible contribution from the systematic errors.\par
Figure~\ref{fig:dist} shows that our measurement, being consistent with previous estimates, has a much smaller error. We attribute this improvement to two main factors: i) the improved data quality, and ii) better estimation and treatment of non-cosmological biases. The former has facilitated the necessity and the possibility of addressing the latter at an improved level. Below we further discuss these aspects.\par
\textit{i) Data quality}: The data used in the latest estimates by~\citett{Bonamente2006, reese2003} are interferometric data  at 30 GHz. The authors estimated the absolute flux calibration at $4\%$ level, and this converts into an $8\%$ error in the final $H_0$ measurement. \textit{Planck} data, instead, have a multi-frequency coverage (in our case six bands at $100-860$ GHz with well-constrained beams) with an error in absolute flux calibration below $0.5\%$ for bands up to $217$ GHz~\citep{Planck4, Planck3}. It spatially resolves and maps clusters near $r_{500}$ , allowing constraints on cluster pressure profile shapes and dispersion. On the analysis side, the \textit{Planck} multi-frequency coverage allows an accurate cleaning of foregrounds and backgrounds for the separation of the SZ signal (see~\citealt{Bourdin}). All of these factors substantially minimise many uncertainties on the Compton parameter $y$ that were present previously as a result of the limited frequency coverage and impossibility of foreground cleaning (e.g. contamination from point sources, kSZ confusion, and radio halos).\par

\textit{ii) Estimation and treatment of biases}: As described before, the currently improved data quality requires a more careful treatment of the biases of non-cosmological origin. In previous works these uncertainties have been estimated independently and were added in quadrature, while we here accounted for correlations between the various sources. The main sources of these uncertainties are the clumpiness of the cluster gas and the departure from spherical shape and from uniform radial thermal distribution (see e.g.~\citealt{reese2010, roettinger, kahawara, wang2006, ameglio, yoshikawa, molnar2002}). We discuss how these effects were considered in the literature in relation to the outcome of our work.\par
The estimates of the clumpiness effect in previous works has been controversial. While there is consensus that clumpiness biases the estimate of $H_0$ high, the size of the effect is still debated.~\cite{kahawara} claimed that the effect is of the order of $10\%-30 \%$. Based on numerical simulations,~\cite{yoshikawa} claimed the effect to be negligible at low redshifts.~\cite{reese2003} and~\cite{Bonamente2006} ignored it for the negligible contribution of clumps that could be detected and excluded from the X-ray data~\citep{LaRoque2006}. On the other hand, more recent simulations seem to consistently point towards a  value close to $1$ near the centre, reaching values significantly higher near $r_{200}$~\citep{nagailau, ronca, vazza, battaglia, planelles}. On average, the clumpiness factor is estimated to be $<1.1$ within $r_{500}$ , which is the radius range relevant for this work. On the observational side, a number of recent works~\citep{morandi2013, eckert2013, eckert} have also started to converge towards similar results.\par
The assumption of spherical symmetry, instead, may bias the $H_0$ estimate in both directions depending on the orientation of the cluster. ~\citett{sulkanen, greinge, udomprasert2004, Bonamente2006} estimated a contribution of $15\%$ for a single cluster, while~\citett{reese2000, reese2002, reese2003} reported it to be around $20\%$. Some authors, under the assumption that the scatter introduced by the asphericity of the clusters will average out for a large sample,  considered their large set exempt from orientation bias~\citep{Bonamente2006, kahawara, sulkanen}. However, in Fig.~\ref{fig:priors} we demonstrate that the contribution from  asphericity is not symmetric and could eventually introduce biases. Our treatment of the $\mathcal{B}$ factor properly accounts for this potential source of error.\par
Finally, isothermality of the gas distribution has usually been assumed, leading to an underestimation of $H_0$ by at least $10\%$~\citep{reese2010, kahawara}, even if other studies quantify the error as high as $10-30\%$~\citep{inagaki, roettinger}.\par
Many sources of errors that were previously considered as systematics are coherently treated in our statistical Bayesian approach thanks to the informed priors. As shown in Fig.~\ref{fig:dist}, our combined treatment of the uncertainties finally results in a $4\%$ overall error accounting for all the above-mentioned effects and for the uncertainties in the values of $\Omega_M$ and $Y$.\par 
It is important to mention that in this work, which relies on \textit{XMM-Newton} and \textit{Planck} observations, we did not account for the cross-calibration issues between \textit{XMM-Newton} and other X-ray instruments. Instead we assumed that the \textit{XMM-Newton} observed temperature represents the true temperature of the ICM, while it is known that \textit{Chandra} returns higher ICM temperatures. Estimates of this temperature discrepancy, however, show a large scatter in the literature (e.g.~\citealt{snowden, nevalainen, martino, schellen}). In the current literature it is easy to find temperature discrepancies of anywhere between $0-20\%$ for $kT \approx 6$keV, which corresponds to the median temperature of our sample. This is partly because temperature measurements strongly depend on the choices of the detectors and/or energy bands. A percentile variation in temperature for a single cluster results in a variation of $\eta_T$ by the same percentile amount in the same direction. Despite this, it is not trivial to estimate the total effect on $H_0$ due to the entire sample because i) the sample has a temperature distribution and the discrepancy depends on the actual cluster temperature; ii) in our analysis we probe external regions ($\approx r_{500}$) of the cluster, where the cluster temperatures drop. This will likely reduce the overall effect. It is beyond the scope of this paper to address this specific issue in detail, and we leave its investigation for future work.\par

\addtocounter{footnote}{-5} 
 \stepcounter{footnote}\footnotetext{Multiple Mirror Telescope Observatory, Arizona, USA}
 \stepcounter{footnote}\footnotetext{The Advanced Satellite for Cosmology and Astrophysics, X-ray astronomy mission, Japan, USA}
 \stepcounter{footnote}\footnotetext{Wilkinson Microwave Anisotropy Probe}
 \stepcounter{footnote}\footnotetext{Millimeter and Infrared Testagrigia Observatory, Val d’Aosta, Italy}
 \stepcounter{footnote}\footnotetext{``Beppo'' Satellite per Astronomia a raggi X, X-ray Satellite, Italy, Holland}
\subsection{Comparison with other cosmological probes}
In this subsection we discuss the comparison of our result with derivations of the value of $H_0$ using other cosmological probes. As described in the introduction, we cannot easily attribute the discrepancy between the high-redshift CMB estimate ($H_0=66.93 \pm 0.62$ km s$^{-1}$ Mpc$^{-1}$~\citealt{Planck1}) and the low-redshift estimate using SNIa ($H_0=73.48 \pm 1.66$ km s$^{-1}$ Mpc$^{-1}$~\citealt{ries2018} or $H_0=73.52 \pm 1.62$ km s$^{-1}$ Mpc$^{-1}$ from~\citealt{ries20181}) to a specific systematic in one particular measurement. In particular, alternative measurements that avoid the use of one or the other data set are either so far unable to reach the given precisions or also encounter similar inconsistencies~\citep{ries2018, bernal}. For example, the measurement using time delays of strongly lensed images of three quasars by the H0LiCOW project~\citep{bonvin} leads to a completely independent measurement of $H_0=71.9^{+2.4}_{-3.0}$ km s$^{-1}$ Mpc$^{-1}$ for a flat $\Lambda$CDM with $H_0$ and $\Omega_{\Lambda}$ left free. When allowing only $H_0$ to vary, the result is instead $H_0=72.8 \pm 2.4$ km s$^{-1}$ Mpc$^{-1}$. Through geometric distance measurements to the megamaser galaxy NGC 5765b,~\cite{gao2016} determined $H_0=66.0\pm6.0$ km s$^{-1}$ Mpc$^{-1}$. Other probes based on the baryon acoustic oscillations (BAO) combined with data independent of CMB and SNIa provide $H_0=67.2^{+1.2}_{-1.0}$ km s$^{-1}$ Mpc$^{-1}$ by~\cite{abbott2017} (BAO and \textit{Dark Energy Survey} Year 1 clustering and weak-lensing data combined with Big Bang nucleosynthesis information), $H_0=66.98 \pm 1.18$ km s$^{-1}$ Mpc$^{-1}$ from~\cite{addison} (BAO from galaxy and Ly$\alpha$ forest with an estimate of primordial deuterium abundance). At the same time, BAO in combination with other CMB measurements from the \textit{Wilkinson Microwave Anisotropy Probe} (\textit{WMAP}), the \textit{South-Pole Telescope} (\textit{SPT}), the \textit{Atacama Cosmology Telescope}(\textit{ACT}), and SNIa data provide $H_0=69.6 \pm 0.7$ km s$^{-1}$ Mpc$^{-1}$ from~\cite{bennet} or in combination with observational Hubble datasets and SNIa data $H_0=69.4 \pm 1.7$ km s$^{-1}$ Mpc$^{-1}$ from~\cite{haridasu}.\par 

In the light of the possible cosmological solutions suggested so far ~\citep{ries, bernal,evslin,lin} that tried to resolve the discrepancy between the CMB and local measurements of $H_0$ , we briefly discuss the importance of these scenarios for our measurement, which is perfectly consistent with the result of~\cite{Planck1} and is compatible within 2$\sigma$ with~\cite{ries2018}. The current possible cosmological extensions aim to decrease the discrepancy between the CMB measurement and low-redshift SNIa measurements by either modifying early-Universe physics (e.g. BBN, density of relativistic species) in order to increase the CMB result or by changing the late-time evolution of the Universe in order to introduce recent evolution in the value of $H_0$ allowing for the difference between the early- and late-time measurements. Given that our result is based on low-redshift data, and it is in the lower range of values, modifications of late-time evolution aiming to bring to an increase in the value of $H_0$ in the local Universe are not required by our findings. The possible modifications in early-time evolution would modify the CMB measurement by moving it towards higher values. In particular, any changes in the value of $Y$ (the only relevant quantity in this work) that were required to reconcile the CMB $H_0$ constraints with the SNIa would lower our measurement of $H_0$.\par 

Thus, being a method that does not rely on additional distance ladders, the use of galaxy clusters to determine a local value of the Hubble constant could be decisive in solving this issue in the future. However, more stringent constraints are required for solid conclusions.\par
\subsection{Estimating possible improvements to accuracy}
\begin{figure}
\centering
\includegraphics[width=8cm]{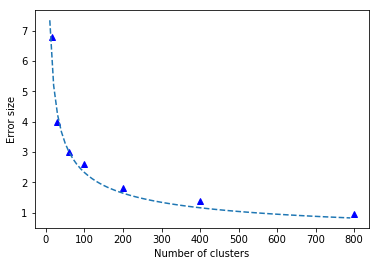}
\caption{Prediction of the error depending on the sample size (blue triangles) over-plotted with the function $3.0\times\left(\frac{60}{N}\right)^{1/2}$ (dashed line).}
\label{fig:errs}
\end{figure}

An improvement to the current result could be achieved by applying this method to larger samples. To estimate the accuracy that could be reached, we created a toy model with ``measurements'' of $\eta_T$ distributed as our sample (with the same mean and scatter). To these data points we assigned errors equal to the average error on $\eta_T$ in our data. We then repeated the procedure to fit the value of $H_0$ for various sizes of our toy sample. In Fig.~\ref{fig:errs} we report our estimated error size for various sample sizes. We found the final error to vary roughly as the square root of the number of clusters ($\frac{1}{\sqrt{N}}$), as demonstrated in the figure with the overlaid line corresponding to $3.0\times\left(\frac{60}{N}\right)^{1/2}$. It is remarkable how a future application of this technique to a sample of $200$ clusters would narrow down the error on $H_0$ to a $3\%$ level, as is also shown by the point added to Fig.~\ref{fig:dist}. High-quality \textit{Chandra} and \textit{XMM-Newton} follow-up observations of the \textit{Planck} cluster catalogue are ongoing and are expected provide us with such constraints in the near future.

\begin{acknowledgements}
We thank Alex Saro and Balakrishna Sandeep Haridasu for useful discussions, and we acknowledge financial contribution from the agreement ASI-INAF n.2017-14-H.O, from ASI Grant 2016-24-H.0., and from “Tor Vergata” Grant “Mission: Sustainability” EnClOS (E81I18000130005)
\end{acknowledgements}

\bibliographystyle{aa} 
\bibliography{bibly}

\onecolumn

\begin{appendix}

\section{Dependence of $\eta_T$ on cluster and cosmological parameters}
\label{app-a}
In this appendix we derive the analytic form of the dependence of $\eta_T$ on the parameters describing the cosmological model and the cluster. We have in total three observable quantities that are projected quantities along the LOS that we list below. \par
\begin{figure}
\centering
\includegraphics[width=8cm]{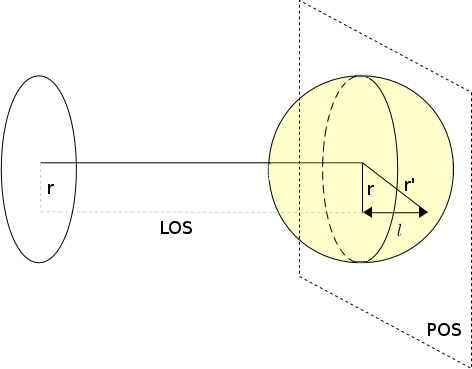}
\caption{Schematic of the signal projection and the definitions used in Appendix~\ref{app-a}. POS and LOS denote the POS and the LOS direction. $r$ is the projected distance from the cluster centre, $r'$ the 3D spherical distance from cluster centre, and $l$ the distance in the direction of LOS from the cluster centre. Given these definitions, we can write $r'^2=r^2+l^2$.}
\label{fig:design}
\end{figure}

1) The SZ signal is proportional to the integral of the pressure of ICM electrons along the LOS. For a given projected distance $r$ on the plane of the sky (POS) from the cluster centre, the signal is
\begin{equation}
y(r)=\frac{\sigma_T}{m_e c^2} \int_{l=-\infty}^{\infty} P_e(l(r))\times dl,
\label{eqn:y-def}
\end{equation}
with $\sigma_T$, $m_e$ , and $P(l)$ being the Thomson scattering cross-section, the mass of electron, and the pressure of electrons in the ICM, respectively, and $l$ being the physical distance along the LOS (see Fig.~\ref{fig:design}). \par

2) The X-ray surface brightness is proportional to the integral of the square of the number density of electrons along the LOS:
\begin{equation}
S_X(r)= \frac{1}{4\pi(1+z)^3}\int_{l=-\infty}^{\infty} [n_pn_e](l(r)) \Lambda(T, \xi, Y) \times dl,
\label{eqn:sx-def1}
\end{equation}
where $z$ is the cluster redshift, $n_p$ and $n_e$ are the number density of protons (i.e. hydrogen, denoted also as $n_H$ in literature) and electrons in the ICM. The factor $\Lambda(T, \xi, Y)$ is the cooling function that depends on the temperature, the abundance of helium (represented here with $Y$), and abundances of other elements heavier than helium in the cluster gas ($\xi$). This function combines the emission due to all the elements responsible for the X-ray continuum emission, mainly hydrogen and helium.  We also note that here the surface brightness has units cnt s$^{-1}$ cm$^{-2}$ arcmin$^{-2}$. If energy units are used, such as erg s$^{-1}$ cm$^{-2}$ arcmin$^{-2}$ , then the factor $1+z$ in the denominator would be to the fourth power, that is, $\Sigma_x(r)=\frac{1}{4 \pi (1+z)^4} \int [n_p n_e](r) \Lambda(T,Z) dl$.\par
We converted $n_pn_e=\frac{n_p}{n_e} \times n_e^2$, leaving $n_e(l(r))^2$ as a profile to be modelled and $\frac{n_p}{n_e}$ a factor that is fixed by the abundance of other elements in the gas relative to hydrogen. At the end of this appendix, we show how this factor is related to the element abundances in the gas. On the other hand, the factor $\Lambda(T, \xi, Y)$ can be expressed as the sum of contributions from hydrogen and helium separately, namely $\Lambda(T, \xi, Y)=\Lambda_H(T)\times (1+4 \frac{n_{He}}{n_p})$, where $\Lambda_H(T)$ denotes the contribution only from hydrogen and the factor $4 \frac{n_{He}}{n_p}$ adds to this the contribution from helium only (assuming all the other elements contribute negligibly to the X-ray continuum). Thus the final form of the X-ray surface brightness is the following:
\begin{equation}
S_X(r)=\frac{1}{4\pi(1+z)^3} \times \frac{n_p}{n_e} (1+4 \frac{n_{He}}{n_p}) \times \int_{l=-\infty}^{\infty} n_e(l(r))^2 \Lambda_H(T) \times dl,
\label{eqn:sx-def}
\end{equation}

 3) The projected temperature, which we approximate with the spectroscopic-like temperature of~\cite{mazz}:
\begin{equation}
kT_{X}(r)=\frac{\int_{l=-\infty}^{\infty} W(l(r)) kT(l(r)) \times dl}{\int_{l=-\infty}^{\infty} W(l(r))\times dl},
\label{eqn:tx-def}
\end{equation}
where $W(l)=\frac{n_e(l)^2}{kT(l)^{3/4}}$, $k$ is the Boltzman constant and $T$ is the temperature of the cluster gas.\par

Next we write the previous equations in function of the projected POS distance from the cluster centre $r$, using the relation $r'^2=r^2+l^2$, where $r'$ is the the 3D spherical distance from the cluster centre. We also convert the quantities representing physical distance to quantities defined relative the characteristic scale of the cluster $x=\frac{r}{r_{500}}$. For the three observables, this conversion would result in the following forms:
\begin{equation}
y(\theta)=P_0e_{LOS} \times D_A \theta_{500}  \times f_{SZ},
  \label{eqn:y-prof-fin}
\end{equation}
\begin{equation}
S_X(\theta)=\frac{1}{4\pi(1+z)^3} \times \frac{n_p}{n_e} (1+4 \frac{n_{He}}{n_p}) \times n_{e0}^2C_{\rho}e_{LOS} \times D_A \theta_{500} \times f_{X},
  \label{eqn:x-prof-fin}
\end{equation}
\begin{equation}
T_{X}(\theta)=kT_0\times f_T,
  \label{eqn:tx-prof-fin}
\end{equation}
where in order to simplify the form of the equations, we have defined the factors 
\begin{eqnarray*}
f_{SZ} \equiv 2\int_{x'=x}^{X_{max}} \frac{(P(x')/P_0) x'}{\sqrt{x'^2-x^2}} \times dx', \\
f_{X} \equiv 2\int_{x'=x}^{X_{max}} \frac{(n_e^2(x')/n_{e0}^2) \Lambda_H x'}{\sqrt{x'^2-x^2}} \times dx',\\
f_{T} \equiv \frac{\int_{x'=x}^{X_{max}} \frac{W(x')kT(x')/kT_0 x'}{\sqrt{x'^2-x^2}} \times dx'}{ \int_{x'=x}^{X_{max}} \frac{W(x') x'}{\sqrt{x'^2-x^2}} \times dx'},
 \end{eqnarray*}
and taken out the normalisations of all quantities $P_e$, $n_e$ , and $T$. We also express $r_{500}$ in terms of the observable quantity $\theta_{500}=\frac{r_{500}}{D_A}$, where $D_A$ is the angular diameter distance to the cluster. The factor $e_{LOS}$ is introduced in all the integrals as a correction for the approximation of spherical symmetry. In the simplest case of an elliptical cluster aligned with any of its axes along the LOS, this quantity could be written as $e_{LOS}=\frac{r_{\parallel}}{r_{\bot}}$, where $r_{\parallel}$ and $r_{\bot}$ are the size of the cluster in the LOS direction and in the POS, respectively. In more complicated cases of random orientation and more complicated shape of the cluster, this factor has a complex dependence on those quantities and can be treated as a general correction term for asphericity. Finally, the factor $C_{\rho}\equiv \frac{<\rho^2>}{<\rho>^2}$ is introduced as a correction taking care of the clumpiness of the cluster gas. The clumpiness results in enhanced X-ray emission, so  that $C_{\rho}$ by definition is always larger than $1$.\par
Equations ~\ref{eqn:y-prof-fin}, ~\ref{eqn:x-prof-fin}, and ~\ref{eqn:tx-prof-fin} exhibit the explicit dependence of the observables on the normalisations of the cluster properties such as $P_e$, $n_e$ , and $T$, the cosmological model, and the deviations of the cluster gas from our assumptions. We next try to relate them to the parameter $\eta_T$ as derived in Sect.~\ref{method}.\par
\vspace{0.3cm}
In order to do this, we distinguish a case $1$) in which  the true cosmological model and parameters as well as the cluster characteristics are known from a case $2$) that corresponds to our modelling described in Section~\ref{method}, which does not necessarily have to be the true model of the Universe and the cluster: a $\Lambda$CDM cosmological model with $\bar{H}_0=70$ km s$^{-1}$ Mpc$^{-1}$, $\bar{\Omega}_m=0.3,$ and $\bar{\Omega}_{\Lambda}=0.7$ and spherical clusters without clumps. Hereafter we use the bar notation to distinguish the parameters for which a value potentially different from truth has been assumed.\par
Our fitting procedure returns $\eta_T$ that in essence reflects the discrepancy of the normalisation of profiles measured with X-ray and SZ observables. Then for case $1$), Eq. \ref{eqn:y-prof-fin}, \ref{eqn:x-prof-fin}, and \ref{eqn:tx-prof-fin} can be combined in the following way:
\begin{equation}
\frac{kT_0}{P_0/n_{e0}}=\left<\frac{f_{SZ}}{y(\theta)} \right>_{\theta}\times \left<\frac{kT_{X}(\theta)}{f_T}\right>_{\theta} \times \left<\sqrt{\frac{S_X(\theta)}{f_X}}\right>_{\theta} \times \left(\frac{1}{n_p/n_e}\right)^{1/2} \left(\frac{1}{1+ 4 \frac{n_{He}}{n_p}}\right)^{1/2} \times \left( \frac{e_{LOS}}{C_{\rho}} \right)^{1/2} \times \theta_{500}^{1/2} D_a^{1/2} (1+z)^{3/2} \equiv 1,
  \label{eqn:real-case}
\end{equation}
where we denote with the brackets $<>_{\theta}$ the fact that the fitted normalisation values carry averaged information from the observed profiles. The equivalence on the right comes from the relation $P_0 \equiv n_{e0} \times kT_0$.\par
For case $2$), the equivalence is instead not true, since our assumptions bias the measured values of $P_0$, $n_{e0}$ , and $T_0$: 
\begin{equation}
\eta_T \equiv \frac{k\bar{T}_0}{\bar{P}_0/\bar{n}_{e0}}=\left<\frac{f_{SZ}}{y(\theta)} \right>_{\theta}\times \left<\frac{kT_{X}(\theta)}{f_T}\right>_{\theta} \times \left<\sqrt{\frac{S_X(\theta)}{f_X}}\right>_{\theta} \times \left(\frac{1}{\bar{n}_p/\bar{n}_e}\right)^{1/2} \left(\frac{1}{1+ 4 \frac{\bar{n}_{He}}{\bar{n}_p}}\right)^{1/2} \times \left( \frac{\bar{e}_{LOS}}{\bar{C}_{\rho}} \right)^{1/2} \times \bar{\theta}_{500}^{1/2} \bar{D}_{a}^{1/2} (1+z)^{3/2} \neq 1.
  \label{eqn:mod-case}
\end{equation}
In this case, the contribution from $\bar{C}_{\rho}$ and $\bar{e}_{LOS}$ has been ignored in our fitting procedure described in Section~\ref{method}, so that these quantities have been considered equal to 1 ($\left( \frac{\bar{e}_{LOS}}{\bar{C}_{\rho}} \right)^{1/2}=1$).\par
Removing the factor $\left( \frac{\bar{C}_{\rho}}{\bar{e}_{LOS}} \right)^{1/2}$ and taking the ratio of Eq. \ref{eqn:real-case} and~\ref{eqn:mod-case} brings us to an expression for the ratio $\eta_T \equiv \frac{k\bar{T}_0}{\bar{P}_0/\bar{n}_{e0}}$ measured in our fitting procedure,
\begin{equation}
\eta_T \equiv b_n\frac{k\bar{T}_0}{\bar{P}_0/\bar{n}_{e0}}=b_n\left( \frac{C_{\rho}}{e_{LOS}} \right)^{1/2} \times \left(\frac{\bar{\theta}_{500}\bar{D}_{a}}{\theta_{500} D_a} \right)^{1/2} \times \left(\frac{{n_p/n_e}}{{\bar{n}_p/\bar{n}_e}}\right)^{1/2} \times \left(\frac{1+ 4 \frac{n_{He}}{n_p}}{1+ 4 \frac{\bar{n}_{He}}{\bar{n}_p}}\right)^{1/2},
\label{eqn:final}
\end{equation}
where as a final step we introduced a factor $b_n$ that represents any other biases that are contained in our fitting procedure that have not yet been included in the equations.\par
At this point, we simplify the two factors $\left(\frac{\bar{\theta}_{500}\bar{D}_{a}}{\theta_{500} D_a} \right)$ and $\left(\frac{{n_p/n_e}}{{\bar{n}_p/\bar{n}_e}}\right)^{1/2} \times \left(\frac{1+ 4 \frac{n_{He}}{n_p}}{1+ 4 \frac{\bar{n}_{He}}{\bar{n}_p}}\right)^{1/2}$ in the following way:\par
\begin{enumerate}
\item $\theta_{500}$ vs $\bar{\theta}_{500}$: $\theta_{500}$ is an observable quantity, which means that $\theta_{500}$ , which the cluster appears to have given the true cosmological model and the cluster distance and size, should be equal to the $\bar{\theta}_{500}$ that we observe.

\item Conversion of $\frac{n_{He}}{n_p}$ and $n_p/n_e$ to $Y$ (mass ratio of helium to hydrogen) dependent factors: We introduce the definitions of abundance in terms of number density and mass. We denote the abundance in terms of number density of a given element relative to hydrogen as $n_i/n_p$ (equivalent to $n_i/n_H$). The abundance in terms of mass, instead, is denoted as $m_i/m_H=A_in_i/(A_Hn_p)=A_in_i/n_p$, where $A_i$ is the atomic number of a given element. In our calculations we make use of measurements of element abundances in cluster gas in terms of number density reported by~\cite{Grevesse} normalised by a constant $0.3$. Thus we have the values of $n_i/n_p$ for elements up to $i=26$. Having the $n_i/n_p$ and the atomic number $A_i$ of each element, the ratio $n_p/n_e$ can be expressed in the following way: 
\begin{equation}
\frac{n_p}{n_e}=\frac{1}{1+\frac{n_{He}}{n_p}A_{He}+\sum_{i=3}^{26} \frac{n_{i}}{n_p}A_i}=(1+2\frac{n_{He}}{n_p}+\xi)^{-1},
\end{equation}
where we introduce the notation $\xi \equiv \sum_{i=3}^{26}  \frac{n_{i}}{n_p}A_i$. Given the metallicity and the abundances we use, this quantity is fixed to $\xi=3.46 \times 10^{-3}.$ \par
At the same time, the mass ratio of helium to hydrogen is
\begin{equation}
Y=\frac{m_{He}}{m_H}=\frac{2\frac{n_{He}}{n_p}A_{He}}{1+2\frac{n_{He}}{n_p}A_{He}+\sum_{i=3}^{26} 2\frac{n_{i}}{n_p}A_i}=\frac{4\frac{n_{He}}{n_p}}{1+4\frac{n_{He}}{n_p}+2\xi}.
\end{equation}
Expressing $\frac{n_{He}}{n_p}$ in terms of $Y$ and substituting in the expression of $n_p/n_e$ , we obtain
\begin{equation}
\left(\frac{n_p}{n_e}\right)^{1/2} \times \left(1+ 4 \frac{n_{He}}{n_p}\right)^{1/2}=\left(\frac{2+4Y\xi}{2-Y+2\xi}\right)^{1/2}
.\end{equation}
Substituting the values of $\bar{n}_p/\bar{n}_e=0.852$ and $\frac{\bar{n}_{He}}{\bar{n}_p}=0.0851$ corresponding to the helium abundance assumed in our analysis:
\begin{equation}
\left(\frac{{n_p/n_e}}{{\bar{n}_p/\bar{n}_e}}\right)^{1/2} \times \left(\frac{1+ 4 \frac{n_{He}}{n_p}}{1+ 4 \frac{\bar{n}_{He}}{\bar{n}_p}}\right)^{1/2}=\left(\frac{2+4Y\xi}{2-Y+2\xi}\right)^{1/2} \times \left(\frac{1}{0.852\times 0.0851}\right)^{1/2}
.\end{equation}

\end{enumerate}

The complete form for $\eta_T$ is then the following:
\begin{equation}
\eta_T=b_n \left( \frac{C_{\rho}}{e_{LOS}} \right)^{1/2} \times \left(\frac{\bar{D}_{a}}{D_a} \right)^{1/2} \times \left(\frac{2+4Y\xi}{1.142\times(2-Y+2\xi)}\right)^{1/2}.
\label{eqn:final-fin}
\end{equation}

\newpage
\section{Treating the correlations within the simulated sample}
\label{app-d}
\begin{figure}
\centering
\includegraphics[width=5.5cm]{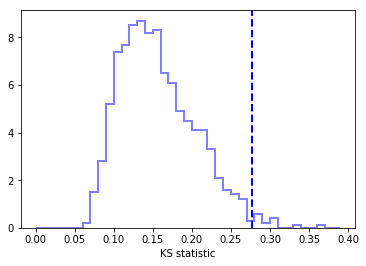}
\caption{Distribution of the Kolmogorov-Smirnov statistic value when comparing the distribution of $\eta_T$ in Fig.~\ref{prior} and the 1000 realisations of the $\eta_T^{(26)}$ distribution for independent cluster sub-sets of the total $216$ cluster set. The dashed line represents the value for which the similarity of the given sub-distribution of $\eta_T^{(26)}$ and the entire distribution $\eta_T$ can be rejected at $0.05$ confidence. These cases constitute only $2\%$ of all. 
}
\label{fig:ks-stats}
\end{figure}

Following the procedure of \cite{rasia2013}, we checked whether using a set of not fully  independent realisations of the same simulated clusters biases our results. We have 26 objects in total, each with three projections at two or three different redshifts: some at $z=0.25$ and $z=0.5$ have masses below our mass cut. This amounts to a total of $216$ realisations. We generated 1000 subsamples of 26 independent realisations by randomly sampling one from the available realisations of each of the 26 objects. We checked through the K-S test the compatibility of the $\eta^{(26)}$ distribution of each of the sub-samples with the overall $\eta_T$ distribution (shown in Fig.~\ref{fig:priors}). We obtained the K-S distribution for the 1000 sub-samples shown in Fig.~\ref{fig:ks-stats}. In only 2 \%\ of the sub-samples can the similarity hypothesis between the distributions (that derived from one sub-sample and the overall distribution) be rejected with a confidence level at least equal to 0.05 (vertical dashed line in the figure).
The figure shows that our final distribution of $\eta_T$ probably does not carry amplified biases that are due to the dependence of the object realisations in the sample. This agrees with the majority of the possible subsets of independent realisations. \par

\newpage
\section{Semi-analytical estimation of bias that is only due to asphericity }
\label{app-e}
Here we give the details of the procedure of deriving a distribution of $\frac{1}{e_{LOS}^{1/2}}$ in a sample similar to our sample of Planck clusters. It can be divided into two main steps: i) given the distribution of SZ mass and redshift of our Planck sample, we derive the corresponding virial mass-redshift distribution from it, and ii) given the virial masses of the clusters, we randomly assign them an elliptical shape in agreement with their mass and an orientation relative to the LOS. We finally project the cluster with given shape and orientation in order to calculate the projection ratio $e_{LOS}=\frac{r_{\parallel}}{r_{\bot}}$, with $r_{\parallel}$ being the size of the cluster along our LOS and $r_{\bot}$ being its projected size on the POS.\par
\subsection{Derivation of $[M_v, z]$ given $[M_{SZ}, z]$}

\begin{itemize}
\item $\pmb{[M_{SZ}, z]}$. We start our calculation starting from the SZ mass-redshift distribution of the \textit{Planck} sample of clusters that we used in this paper. 
\item \textbf{Derivation of} $\pmb{[M_{500}, z]}$. Taking an approximate value of the SZ bias for \textit{Planck} clusters as $b_{SZ}=0.25$ (see \citealt{Plcluster2015, sereno2017}), we derive $M_{500}=(1+b_{SZ})M_{SZ}$.
\item \textbf{Derivation of} $\pmb{[M_{200}, c_{200}, z]}$\footnote{$M_{\Delta_x}$ is defined as the mass of the galaxy cluster at a radius at which the density of the cluster is $\Delta_x$ times the critical density of the Universe at a given redshift. Similarly, $c_{\Delta_x}$ is defined as the concentration at the same radius.}. \cite{hukravtsov} provide a formula for conversions between definitions of halo mass under the assumption of a Navarro-Frenk-White (NFW)\ density profile. Thus given a set $[M_{\Delta_x}, c_{\Delta_x}, \Delta_x]$, where $M_{\Delta_x}$ and $c_{\Delta_x}$ are the known mass and concentration parameter at a given mass overdensity $\Delta_x$  , they provide a formula to derive a set of $[M_{\Delta_y}, c_{\Delta_y}]$ at overdensity $\Delta_y$. At the same time, \cite{meneghetti} provide a formula to relate $c_{200}$ to $M_{200}$ in the following form: $c_{200}(M_{200}, z)=A \times \left( \frac{1.35}{1+z} \right) ^B \times \left( \frac{M_{200}}{8 \times 10^{14} h^{-1} M_{\odot}} \right) ^C$, where the values of parameters $A$, $B$ and $C$ are fit assuming a NFW density profile using the MUSIC 2 hydrodynamic simulated massive clusters. The relationship also includes an intrinsic scatter. Using these conversions and the scatter in agreement with~\cite{meneghetti} , we can establish the pair $[M_{200}, c_{200}=c_{200}(M_{200}, z)]$ that corresponds to each pair of $[M_{500}, z]$ in our set.\par
\item \textbf{Derivation of} $\pmb{\Delta_v}$\footnote{Overdensity at the virial radius relative to the critical density of the Universe at a given redshift.} \textbf{and} $\pmb{[M_v, z]}$. \cite{bryan} derived an approximate scaling formulae for a number of virial quantities for a range of redshifts in three different cosmological models. In particular, they fit a formula for the overdensity at virial radius relative to the critical density of the Universe given the redshift and the cosmological model. Converting their formula to derive the overdensity relative to the mean matter density, we derive a value of $\Delta_v$ for the redshift of each cluster. Then given $[M_{200}, c_{200}, \Delta_{200}, \Delta_v]$ again referring to \cite{hukravtsov}, we derive a value of $M_v$ for each cluster in our set.
\end{itemize}
At the end of this first part, we then have the distribution of $[M_v, z]$ of our \textit{Planck} sample of clusters. We can now create a large representation of this distribution. We pull out $N=30000$ realisations of $M_v$ from this distribution.

\subsection{Derivation of the asphericity ratio distribution given $M_v$}

In this part we use the $N$ values of the virial mass and chose random shapes and orientations in agreement with these masses in order to calculate the projection ratio $e_{LOS}$ for these $N$ clusters. We later assume that this represents the distribution of $e_{LOS}$ of our set of clusters.
\begin{itemize}
\item \textbf{DM axis ratios}. Using Millennium XXL simulations,~\cite{bonamigo2014} derived simple functional forms for axis ratio distributions of clusters with given virial mass. These formulae take as input the cluster virial mass and output a probability distribution function for the  minor-to-major and intermediate-to-major axis ratios. Given the $N$ masses of clusters of our set, we chose $N$ random values for the axis ratios following these distributions.
\item \textbf{ICM axis ratios}. \cite{kawahara2010} constructed a model that allows deriving the axis ratio of the gas distribution in the cluster based on the DM distribution and the assumption of hydrostatic equilibrium (HE). This is done based on the argument that the matter distribution follows the DM isopotentials due to HE. Following \cite{kawahara2010}, we calculate the isopotential surfaces for a given DM distribution. Considering the cluster gas to follow the isopotentials, we take the ratio of axes of these surfaces as axis ratios of the gas distribution. 
\item  \textbf{Random orientation with respect to the LOS}. We chose random orientations for the $N$ clusters using the $\text{three}$ Euler angles.
\item \textbf{Projection of the constructed ICM ellipsoid to derive $\pmb{e_{LOS}=\frac{r_{\parallel}}{r_{\bot}}}$ and consequently $\pmb{\frac{1}{e_{LOS}^{1/2}}}$}. \cite{sereno2017} provided formulae for projecting a triaxial ellipsoid with a given orientation defined by the $\text{three}$ Euler angles. The eventual derived quantity is the ratio of the LOS length of the cluster and the size of the cluster projection in the POS - $e_{LOS}=\frac{r_{\parallel}}{r_{\bot}}$. We project the constructed ellipsoid in the POS and the LOS following \cite{sereno2017}. As a measure for $r_{\bot}$ , we take the geometrical average of the two axes of the cluster projection on the POS. The exact equations for calculating these quantities are given in \cite{sereno2017}. 
\end{itemize}
In conclusion, we use our distribution $[M_{SZ}, z]$ in order to derive a distribution of $\frac{1}{e_{LOS}^{1/2}}$. The final distribution is shown in Fig.~\ref{fig:priors}.

\newpage
\section{Implementing a tabulated distribution as a prior}
\label{app-c}

\begin{figure}
\centering
\includegraphics[width=8cm]{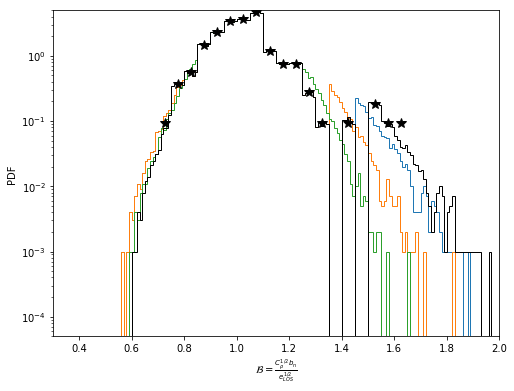}
\includegraphics[width=9.5cm]{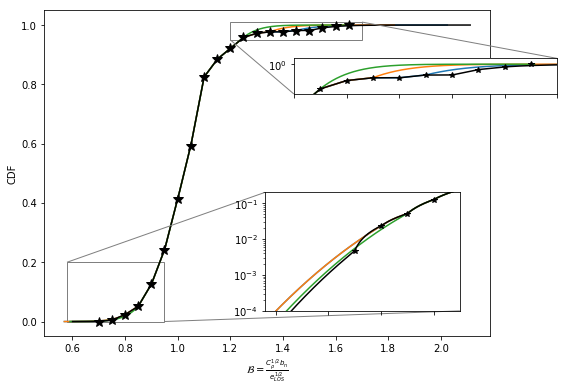}
\caption{Extended probability (left) and cumulative (right) distribution functions of $\mathcal{B}$ using 1, 3, 5, and 7 counts for the extension shown as black, blue, orange, and green lines, respectively. The actual points of the distribution resulting from the $216$ values derived using simulations are shown with stars.}
\label{fig:pdf_prior}
\end{figure}

In this section we describe the extrapolation of the distribution of $\mathcal{B}$ derived in Section~\ref{prior} over the range explored during the sampling. In Fig.~\ref{fig:pdf_prior} we show the form of the probability and the cumulative distributions we derived. The values corresponding to each bin are shown with stars. Given our set of $216$ clusters, this distribution is defined only over the range $0.7<\mathcal{B}<1.65$. We need to extrapolate it to a wider range of values in order to use it as our prior $p(\mathcal{B})$.\par 
In order to do this, we chose the next to last bins from the tails of the distribution and extended the distribution starting from these points assuming a Gaussian (black line in the figures). In the probability distribution function on the left, we basically ignored the last bins of the histogram tail and extended the area spanned by these bins under a Gaussian shape.\par
We achieved this by means of the cumulative distribution function (CDF), since the CDF naturally takes care of the normalisation of the final distribution. We cut the CDF at given bins and extended from that point on with a Gaussian CDF with a mean equal to 1.03 (as for our distribution) and with a standard deviation such that the Gaussian CDF passes exactly through the required bin value. The final CDF was then composed of the analytic tails approximated by us and of the tabulated central part that is the direct result of the simulations.\par
We note that the left side of the resulting distribution agrees well with a Gaussian continuation. The right side, however, seems to be flatter. In order to check the validity of our extrapolation, we repeated the same exercise, but started the extrapolation from bins farther away from the tails, ignoring enough bins to skip 3, 5, or 7 counts from the tails. The curves shown for these extrapolations (blue, orange, and green, respectively) prove once again that the approximation for the Gaussian is good enough for the left side of the distribution, but not for the right side. Despite this small disagreement, we note that the estimated value of $H_0$ is well within the errors for the extrapolations that do not completely remove the flatness of the right tail (using the one- and three-point approach). 

\end{appendix}
\end{document}